\documentclass[a4paper,fleqn,usenatbib]{mnras}
\pdfoutput=1
\usepackage{newtxtext,newtxmath}

\usepackage[T1]{fontenc}
\usepackage{ae,aecompl}

\usepackage{graphicx}	
\usepackage{amsmath}	
\usepackage{amssymb}	
\usepackage[utf8]{inputenc}
\usepackage{bm}

\newcommand{\pd}{\partial}

\def\apjl{ApJL}

\def\aap{A\&A}
\def\mnras{MNRAS}
\def\araa{ARA\&A}

\def\Rs{R_{\odot}}

\newcommand{\brac}[1]{\langle #1 \rangle}
\newcommand{\mean}[1]{\overline{#1}}

\def\be{\begin{equation}}
\def\ee{\end{equation}}
\def\ba{\begin{eqnarray}}
\def\ea{\end{eqnarray}}

\def\alf{Alfv\`en }
\def\bvf{Brunt-Vais\"ala }

\title[Tayler instability in stellar interiors]{Global simulations of Tayler instability in stellar interiors:\\ The stabilizing effect of gravity}

\author[G. Guerrero et al. ]{G. Guerrero, $^{1}$\thanks{E-mail: guerrero@fisica.ufmg.br}
 F. Del Sordo$^{2,3}$  \thanks{E-mail:fabiods@ia.forth.gr}
 A. Bonanno $^{4}$ 
 P.~K. Smolarkiewicz $^{5}$
\\
$^{1}$Physics Department, Universidade Federal de Minas Gerais,
Av. Antonio Carlos, 6627, Belo Horizonte, MG 31270-901, Brazil  \\
$^{2}$Institute of Astrophysics, FORTH, GR-71110 Heraklion, Greece  \\
$^{3}$Department of Physics, University of Crete, GR-70013 Heraklion, Greece \\
$^{4}$INAF, Osservatorio Astrofisico di Catania, via S. Sofia, 78, 95123 Catania, Italy\\
$^{5}$ European Centre for Medium-Range Weather Forecasts, Reading RG2 9AX, UK
}

\date{Accepted 2019 October 4. Received 2019 October 3; in original form 2019 September 12}

\pubyear{2015}

\begin{document}
\label{firstpage}
\pagerange{\pageref{firstpage}--\pageref{lastpage}}
\maketitle

\begin{abstract}
Unveiling the evolution of toroidal field instability, known as Tayler instability, 
is essential to understand the strength and topology of the magnetic fields
observed in early-type stars, in the core of the red giants, or in any stellar radiative zone.
We want to study the non-linear evolution of the instability of a toroidal field stored in a
stably stratified 
layer,
in spherical symmetry 
and
in the absence of rotation.
In particular, we intend to quantify the suppression of the instability
as a function of the Brunt-Väisäla ($\omega_{\rm BV}$)  and the Alfvén ($\omega_{\rm A}$) frequencies.
We use the MHD equations as implemented in the anelastic approximation in the EULAG-MHD code and perform
a large series of numerical simulations of the instability exploring the parameter
space for the $\omega_{\rm BV}$ and $\omega_{\rm A}$.
We show that beyond a critical value gravity strongly suppress the instability, in agreement with the linear analysis.
The intensity of the initial field also plays an important role: weaker fields show 
much slower growth rates.
Moreover, in the case of very low gravity, the fastest growing modes
have a large characteristic radial scale, at variance with the
case of strong gravity, where 
the instability is characterized by horizontal displacements.
Our results illustrate that the anelastic approximation can efficiently describe the evolution 
of toroidal field instability in 
stellar interiors. 
The suppression of the 
instability as a consequence of increasing values of $\omega_{\rm BV}$
might play a role to explain the magnetic desert in Ap/Bp stars, since weak fields are only 
marginally unstable in the case of strong gravity.
\end{abstract}

\begin{keywords}
magnetic fields -- stellar evolution -- numerical simulations
\end{keywords}



\section{Introduction}

Recent high quality data from stellar observations have allowed to measure and 
characterize the magnetic field in stars of almost all types 
\citep[see reviews by][]{dola09,Berdyugina09,Mathys12,Ferrario18}.
These observations impose serious challenges to the theoretical models suited to explain such fields.
The turbulent dynamo theory, canonical model for stars with radiative cores and 
convective envelopes,  may be applied to solar-type stars. 
However, dynamo types different to the $\alpha\Omega$ model have to be invoked to explain the fields measured in fully 
convective stars. 

More problematic is the case of main-sequence peculiar A and B type stars 
(so called Ap/Bp stars),  with masses between 1.5 and 6$M_{\odot}$,
representing about $\sim 7\%$ of the A-star population.
The structure of these objects is mostly radiative, lacking a highly turbulent environment appropriate for the 
dynamo to operate.  
Nevertheless, they are characterized by magnetic fields of strengths between $\sim 300$ and $10^3$ G 
(similar numbers have been reported for massive O and B stars and also for pre-main sequence Herbig 
Ae/Be objects). The lack of 
A stars with fields within the $1$ to $300$ G range has been called the 
``Ap/Bp magnetic desert'' \citep{Auriere+07}.

Ap/Bp stars are statistically slower rotators
than other A/B stars and 
the observed magnetic field topology appears rather simple 
when compared to low-mass main sequence stars, yet
no clear correlation with fundamental stellar parameters 
has been found \citep{dola09}.

A possible explanation 
for the origin of this type of magnetism is the fossil-field hypothesis. 
According to this idea
the field originates from the magnetic field in the interstellar medium which gets 
subsequently amplified by compression during the collapse phase of a star. 
A series of numerical simulations \citep{bra08,iba15,dubama10} pionereed by \cite{bn06} have 
shown that a random initial seed field can indeed evolve 
into a topological configuration of mixed, toroidal and poloidal, field components 
with comparable energy 
and stable over several \alf travel times. 
On the other hand numerical and analytical considerations in cylindrical 
geometry suggest that magnetic configurations of the mixed-type can still be 
prone to very high 
longitudinal
mode number $m\gg1$ resonant MHD instabilities. 

The fossil-field hypothesis has been criticized on the basis that if the observed 
field is a relic of the interstellar field from which the star formed, then one would 
expect stars forming in different regions having diverse incidence of magnetism.  
However, this scenario is not  supported by observations \citep{Paunzen+05}. 
Another puzzling  observational fact is the scarcity of close binaries among the population
of main-sequence intermediate-mass magnetic stars. 
For these reasons \cite{fe09} have proposed that the initial field configuration 
might be a toroidal magnetic field resulting from the strong differential rotation 
produced by merger events.  
In turn, the toroidal field configuration may either remain stable hidden in 
deeper layers, or decay due to Tayler-like instability into a stable configuration of 
mixed fields. In both cases the field will decay afterwards on diffusive time 
scales.  

MHD instabilities in stable stratified stellar plasmas 
might also play 
a central role in the transport of angular momentum in radiative zones, 
explaining
the slow rotation of the core of the red giants \citep{be12,triana}, 
the suppression of the dipolar mixed modes in the core of the red 
giants \citep{fu15}, and as source of an $\alpha$-effect in the solar 
tacochline \citep{sule,gg18}. 
From linear analysis we have learnt that
rotation plays a stabilizing role \citep{pitts,bu13apj} while thermal diffusivity 
tends to oppose to the stabilizing role of gravity, and the resulting growth rates 
are of the order of the evolutionary time scales according to \cite{bu12apj}.

The use of direct numerical simulations 
to determine stable field configurations has to be properly motivated, as the 
choice of the basic state can play an essential role 
in the growth rate and the non-linear evolution of the instabilities. 
As a matter of fact, 
by construction numerical simulations can only provide sufficient conditions for 
instability to occur, while in general one is interested in knowing the set of necessary 
conditions for stability corresponding to the physical situations at hand.

The first work aiming to encode the evolution of an initially unstable 
toroidal magnetic field in a realistic basic state, including gravity and
differential rotation, was presented by \cite{arlt13}.
They concluded that the 
observed magnetism of Ap stars should be interpreted as a relic of the 
Tayler instability \cite{tayler}.
However, at variance with physical intuition, the authors did not 
detect any stabilizing effect due to gravity in their simulations.

\cite{ga15} 
discussed
the instabilities of a toroidal field created by the winding-up
of an initial poloidal field in a differentially rotating stellar interior. 
They explored the role of the density stratification and tested
different initial conditions in 2D numerical simulations in spherical
geometry.
From 3D numerical simulations of a kinematically-generated toroidal field,
\cite{jo15} proposed the idea that the magnetorotational instability (MRI)
is more efficient than the Tayler instability, at variance with the results by \cite{arlt13}.

In this work we aim to
clarify the role of the initial conditions on the stability properties of
a toroidal magnetic field in a stably stratified plasma. 
We concentrate on non-rotating models which can be fair 
approximation for very slow-rotating systems.
In particular, we focus on the combined role of gravity and the 
initial magnetic field strength
in the development of the Tayler
instability and its subsequent non-linear phase. 
In fact, in a stably stratified plasma, buoyancy 
has a stabilizing effect along the radial direction and the Tayler instability should therefore develop 
along horizontal displacements. On the other hand, in a realistic stellar interior
gravity decreases 
with radius, but in the outer, low-density, regions near the surface the Lorentz 
force is expected to be the leading restoring force which can destabilize a locally 
stored magnetic field. 
In this case,
both the radial and the longitudinal components are 
expected to determine the stability properties of the plasma.

We perform anelastic
global numerical simulations with the EULAG-MHD code.
It is an extension of the hydrodynamic model EULAG
predominantly used in atmospheric and climate research \citep{Prusa08}.
It has been extensively tested in various numerical simulations
of stellar interiors \citep[e.g.,][]{GCS10,Zaire+17,gg18}, 
but never used for a focused study of Tayler instability in stably stratified interiors. 
We will show that EULAG-MHD reproduces the development 
of the instability in agreement with the linear analysis, and it is able to 
follow the further evolution during the non-linear phase. 

At variance with the results presented in \cite{arlt13} we shall show that 
not only the ratio between the local \bvf frequency and the local \alf frequency
determines the onset of the instability, but in general
different radial profiles of the magnetic field might have different 
stability properties in the star interior. 

The structure of the paper is the following: in Section \ref{Sec:stab} we present some 
stability consideration of the problem; in Section \ref{Sec:3D} we discuss our numerical 
approach to the problem that allows us to obtain the results presented in Section
\ref{Sec:res};  in Section \ref{Sec:conc} we draw some conclusions and outline follow up plans.

\begin{figure}
\begin{center}
\includegraphics[width=0.5\textwidth]{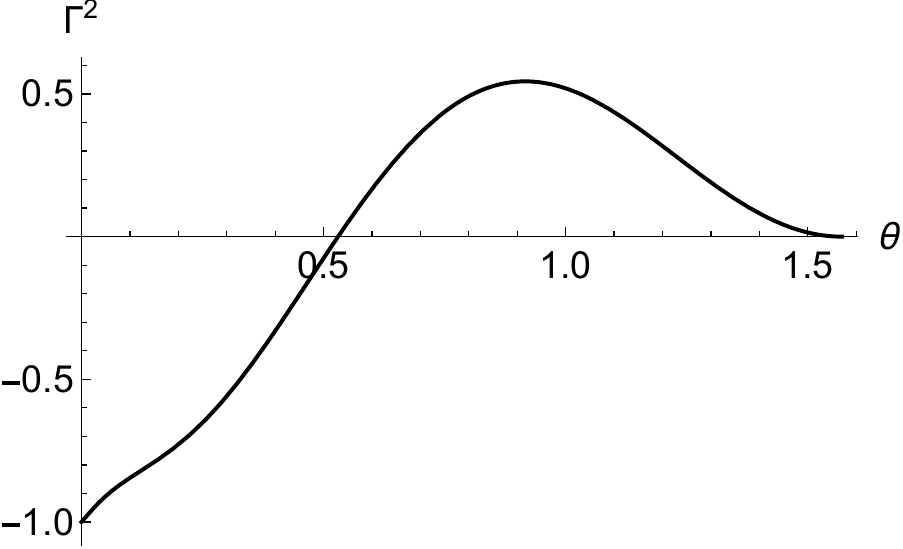}
\caption{Growth rate as a function of $\theta$ for $\delta=0$, for $B_{\varphi}\propto \sin\theta \cos\theta$.  \label{f2}}
\end{center}
\end{figure}

\section{Stability considerations}\label{Sec:stab}
The stabilizing influence of gravity has been recognized since the seminal 
paper by \cite{tayler}. 
In cylindrical symmetry and in absence of vertical field as well as density stratification, 
a necessary and sufficient condition for
the $m=\pm 1$ modes
to be stable,
is
\be
\label{sta}
g_s \frac{\partial \rho}{\partial s} -\frac{\rho^2 g_s^2}{\gamma p} - 
\frac{B^2}{s^2}-\frac{2 B}{s}\frac{d B}{d s} >0 \;,
\ee
where $s$ is the cylindrical radius, $g_s<0$ is the local gravity in 
the $s$ direction, $B$ is the toroidal component of the field, $p$ the pressure of the fluid
and $\gamma$ the adiabatic index. 
For a spherical Couette flow the above 
condition reduces to to 
\be
\frac{d}{ds} (s B^2)<0
\ee
which implies that marginal stability is achieved if the field decreases 
no slower than $1/s^{1/2}$.
It is instructive to rewrite relation (\ref{sta}) in terms of the \bvf 
frequency squared, 
$N^2=g_s(1/\rho \; \partial \rho/ \partial s - 1/\gamma p \, \partial p/\partial s)$~, as
\be
N^2 -\frac{g_s}{\gamma p}\left (\frac{B^2}{s}+B\frac{d B}{ds} \right ) - 
\frac{B^2}{s^2\rho}-\frac{2 B}{s\rho} \frac{d B}{ds}>0 \;,
\ee
from where one can notice that,
if real buoyancy frequency  N
is sufficiently large, the left hand side can 
be positive.

In spherical symmetry the situation is much more involved. 
The first attempt to address the spherical symmetry can be found in 
\cite{g81} using a WKB approximation in radius.
The role  of rotation has been discussed in  
\cite{ki08} and subsequently in \cite{kiru08}, where it was shown 
that the instability is essentially three-dimensional,
as also confirmed by \cite{bour13}.
The specific role of gravity has been discussed in detail in 
\cite{bu12apj}.

In particular, the MHD stability of a  longitudinally uniform toroidal field $B_{\varphi}= B_0 \psi(r, \theta)$
has been studied, in the incompressible limit,  
assuming perturbations of the type $\exp{(\sigma t - i \ell \theta - i m \varphi)}$  with $\ell \gg 1$; 
here $\psi$ is a scalar function, $\sigma$ is the inverse of a typical time scale,
$m$ is the 
longitudinal
wavenumber and $\ell$ the 
latitudinal
wavenumber.
In the limit of vanishing thermal diffusivity, disturbances about the equilibrium configuration 
have been discussed in terms of the normalized growth rate $\Gamma$ governed by the second order differential equation 
\citep[see Eq. 6 in][]{bu12apj}

\ba
\label{stab}
&&\Bigl (\Gamma^2+\frac{m^2\psi^2}{r^2\sin^2\theta}\Bigr)\frac{d^2v_{1r}}{dr^2}
+\Bigl(\frac{4\Gamma^2}{r}+\frac{2}{a}\frac{m^2\psi^2}{r^2\sin^2\theta}\Bigr)\frac{dv_{1r}}{dr}\nonumber\\
&&+\Bigl ( \frac{2\Gamma^2}{r^2}-\frac{m^2}{r^2\sin^2\theta}\bigl(1+\frac{\ell^2}{m^2}\sin^2\theta\bigr) 
\bigl ( \Gamma^2+\delta^2+ \frac{m^2 \psi^2}{r^2 \sin^2\theta} \bigr )  \nonumber\\
&& + \frac{2}{r} \frac{m^2 \psi^2}{r^2 \sin^2\theta}  
\Bigr [ \frac{1}{a} (1+\frac{\ell^2}{m^2}\sin^2\theta )
-\frac{2}{r}\frac{\Gamma^2}{\Gamma^2+\frac{m^2\psi^2}{r^2\sin^2\theta}} \Bigr ] \Bigr) \; v_{1r}=0~,
\label{pippo}
\ea
where $v_{1r}$ is the velocity perturbation along the radial direction r, 
$\Gamma=\sigma / \omega_{A0} $,
is the normalized growth rate, $1/a = \partial (rB_{\varphi}) / \partial r$, and 
\be
\delta^2 = \omega_{BV}^2/\omega_{A0}^2
\ee 
Here we defined
\be
\omega_{A0}^2  = {B_{0}^2}/{4 \pi \rho R^2} \;\;,\;\; \omega_{BV}^2 = - g \beta(\nabla_{ad} T - \nabla T)/T \;,
\label{eq.ndq}
\ee
\noindent
as, respectively, the Alfvén frequency calculated at r=R, and the \bvf frequency, with $\beta$ the thermal expansion 
coefficient.

The growth rate in general depends on $\delta$ and $\theta$.
At large $\delta$, one observes that $\Gamma^2\rightarrow -1$ and the system is always stable: as a consequence there exists
a critical $\delta$ that stabilizes the toroidal field. 
Moreover, there also exist angular regions where the instability is more efficient.
In Fig.~\ref{f2} the angular dependence of the growth rate
is depicted for $B_{\varphi}\propto \sin 2 \theta$. In this case   
the maximum growth rate is at $\theta \approx \pi/4$. 

In the presence of a non-zero diffusivity, the growth rate is never completely suppressed, although 
its characteristic time scale is much greater than the \alf time scale.
It is also known that atomic diffusion may induce instabilities 
\citep[e.g. see][for hydrodynamical instabilities induced in A stars]{Deal2016}
and we expect that this kind of diffusion might play a role by making the magnetic 
fields less stable. 
In the simulations presented in the following sections, we expect to see
a clear suppression of the instability for large $\delta$.  However, due  to
the Newtonian cooling term and to finite numerical thermal diffusivity, in
practice, the suppression will never be complete. Remarkably, the qualitative 
behaviour of the instability can be captured by eq.~(\ref{stab})
consistently with the outcomes of numerical simulations, as we shall see in Sec. \ref{Sec:res}.

\begin{figure}
\begin{center}
\includegraphics[width=0.9\columnwidth]{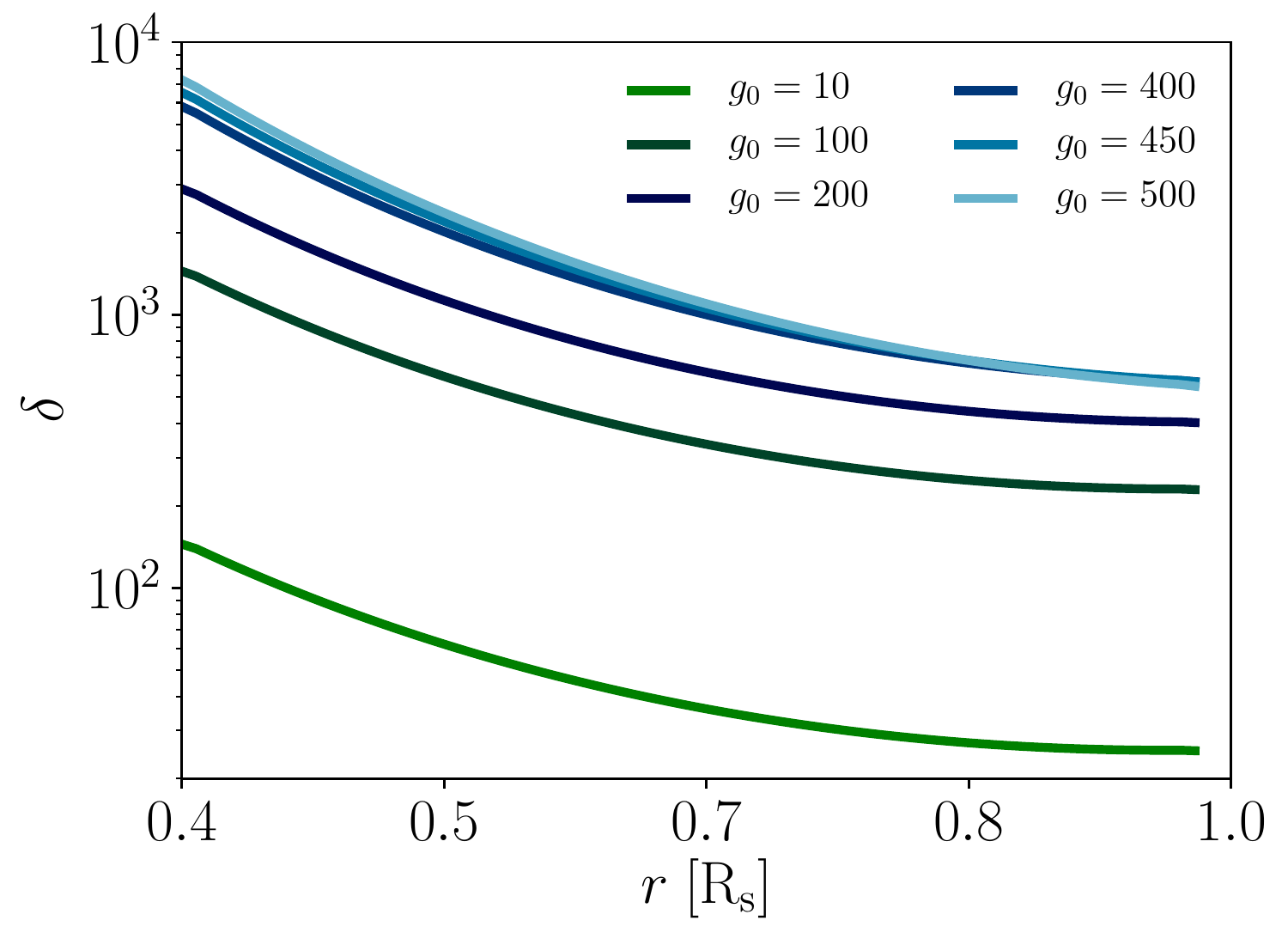}
\caption{Radial profile of $\delta$ for simulations 
TiA04, TiA07, TiA09, TiA12, TiA13 and TiA14. These simulations start with
the same initial magnetic field, $B_0 = 1$ T, but have different gravity at
the bottom of the domain, $g_0$.}
\label{f.delta}
\end{center}
\end{figure}

\begin{table*}
\centering
\begin{tabular}{lccccccccc} 
 Model & $g_0$ & $B_0$ & $\mean{\omega}_{BV} \cdot10^{-8}$ & $\mean{\omega_A}\cdot10^{-7}$ & $\rho_b/\rho_t$ 
       & $\mean{\delta}$ & $B_p/B_t $ & $\Gamma$ & $k_r^{max}$ \\
 \hline
 \hline
TiA00    & 0.01 & 1.00  &  1.3567 &  2.4225  &  1.00  &     0.056   &   0.39   &  5.388  & 1.3    \\
TiA01    &  1     & 1.00  &  136.67 &  2.4239  &  1.01  &     5.638   &   0.46   &  4.531  & 3.3  \\      
TiA02    & 5      &  1.00 &  683.34 &  2.4305  &  1.02  &   28.115   &   0.33   &  3.949  & 10.2  \\
TiA03    & 10    & 1.00  &  1366.7 &  2.4384  &  1.03  &    56.049  &   0.17   &  3.363  & 12.6   \\
TiA04    & 30    & 1.00  &  4100.2 &  2.4716  &  1.11  &   165.890 &   0.25   &  1.573  & 16.1    \\
TiA05    & 50    & 1.00  &  6833.8 &  2.5064  &  1.18  &   272.656 &   0.28   &  0.887  & 19.5    \\
TiA05rf  & 50    & 1.00  &  6833.8 &  2.5064  &  1.18  &   272.656 &   0.32   &  1.071  & 15.3    \\
TiA06    & 70    & 1.00  &  9567.5 &  2.5421  &  1.27  &   376.360 &   0.36   &  0.757   & 17.0   \\
TiA07    &100   & 1.00  & 13668   &  2.5991  &  1.40  &   525.882 &   0.28   &  0.454   & 18.5    \\
TiA08    &150   & 1.00  &  20503  &  2.7024  &  1.66  &   758.719 &   0.32   &  0.345   & 20.2    \\
TiA09    &200   & 1.00  &  27339  &  2.8184  &  1.97  &   970.048 &   0.36   &  0.223   & 16.7    \\
TiA10    &250   & 1.00  &  34176  &  2.9499  &  2.33  & 1158.546 &   0.33   &  0.178   & 18.7     \\
TiA11    &300   & 1.00  &  41014  &  3.1010  &  2.76  & 1322.613 &   0.17   &  0.202   & 22.3     \\
TiA12    &400   & 1.00  &  54691  &  3.4857  & 3.87   & 1569.008 &   0.57   &  0.168   & 18.7     \\
TiA13    &450   & 1.00  &  61531  &  3.7394  & 4.58   & 1645.477 &   0.39   &  0.174   & 16.1     \\
TiA14    &500   & 1.00  &  68372  &  4.0576  & 5.42   & 1685.022 &   0.70   &  0.163   & 15.4     \\
\hline
\hline
TiAhr00   & 0.01&     1.00  & 1.3567  &  2.4206    & 1.00 &    0.056    &  0.57   &    6.542  & 0.98 \\ 
TiAhr01   &    1  &     1.00  & 135.67  &  2.4222    & 1.01 &   5.601    &  0.64  &     4.571 & 5.49\\
TiAhr02   &   20 &     1.00  & 2713.4 &  2.4533    & 1.07 &  110.603    &  0.14  &     3.258 & 27.47\\
TiAhr05   &   50 &     1.00  & 6783.6 & 2.5047    & 1.18 &  270.834     &  -  &     1.716 &41.00\\
TiAhr14   & 500 &     1.00  & 67859 &  4.0616    & 5.49 &  1670.744     &  -  &      0.242 &38.18\\
\hline
\end{tabular} 
\caption{Parameters and results of simulations with fixed magnetic field and different
values of 
$g_0$. 
The table presents the mean
Brunt-Vaisala (buoyancy)
$\mean{\omega}_{BV}$,  and 
Alfvén, $\mean{\omega_A}$ frequencies, the bottom-to-top density contrast $\rho_b/\rho_t$, the parameter 
$\mean{\delta}$, the ratio of 
domain-averaged poloidal and toroidal fields, $B_p/B_t $,
at the end of the linear phase of the 
instability, 
the growth rate $\Gamma$, and the maximum vertical wave number, $k_r^{max}$. 
The lowest rows show results from simulations with higher resolution, double of all the others. 
We report the value of $B_p/B_t $ only for the simulations that
reached the end of the linear growth.
\label{tab.1}}
\end{table*}

\begin{table*}
\centering
\begin{tabular}{lccccccccc} 
Model   & $g_0$  & $B_0$    & $\mean{\omega}_{BV} \cdot10^{-8}$ & $\mean{\omega}_A\cdot10^{-7}$ & $\rho_b/\rho_t$ & $\mean{\delta}$   & $B_p/B_t $& $\Gamma$ & $k_r^{max}$\\
 \hline
 \hline
TiB50a   & 50  &   0.01 &   6833.8    &   2.5064      &   1.18     &   27265.642  &   -  &   0.174 & -\\
TiB50b   & 50  &   0.05 &   6833.8    &   12.532     &   1.18     &   5453.122    &   -  &   0.199 & 17.3\\
TiB50c   & 50  &   0.10 &   6833.8    &   25.064     &   1.18     &   2726.574    &   0.10  &   0.268 & 17.4  \\ 
TiB50d   & 50  &   0.50 &   6833.8   &  125.32    &   1.18     &    545.313       &   0.25  &   0.630 & 18.4\\
TiB50e   & 50  &   0.65 &   6833.8   &   162.91    &   1.18     &    419.471      &   0.28  &   0.771 & 16.5  \\
TiA05    & 50  &   1.00 &   6833.8   &   250.62    &   1.18     &    272.679      &   0.36  &   0.887 & 19.3\\

\hline
TiB100a   & 100  &   0.01  & 13668  &  0.0260    &  1.40   & 52588.239   &  -  &  0.011  $^*$ & -\\
TiB100b   & 100  &   0.05  & 13668  &  0.1299   &  1.40   & 10517.648   &   -   &  0.044       &6.8 \\
TiB100c   & 100  &   0.10  & 13668  &  0.2599   &  1.40   & 5258.824    &   -   &  0.134       & 7.4\\
TiB100d   & 100  &   0.50  & 13668  &  1.2996  &  1.40   & 1051.765    &    0.25    &  0.332       &17.7 \\ 
TiB100e   & 100  &   0.65  & 13668  &  1.6894  &  1.40   & 809.050     &    0.30    &  0.406       & 18.9 \\
TiA07      & 100  &   1.00  & 13668   &  2.5991  &  1.40  &   525.882 &     0.28    &  0.454       & 18.5\\

\hline
TiB150a  &  150   &  0.10   &  20503 & 0.0270   &  1.66  & 75871.9   &      -      &  0.010$^*$ & -\\ 
TiB150b  &  150   &    0.05  &  20503  & 0.1351 &  1.66  & 15174.38  &      -      &  0.032$^*$ & -\\ 
TiB150c  &  150   &  0.10   &  20503 &  0.2702  &  1.66  &  7587.189  &     -      &  0.039  &6.8  \\
TiB150d  &  150   &  0.50  &  20503  & 1.3512  &  1.66  &  1517.44  &       0.10       &  0.226 & 16.4\\
TiB150e  &  150   &  0.65  &  20503  & 1.7565  &  1.66  &  1167.28  &       0.26       &  0.285 & 17.2\\
TiA08     &   150   & 1.00  &  20503  &  2.7024  &  1.66  &   758.719 &     0.32    &  0.345   & 20.2 \\
\hline
\end{tabular} 
\caption{Results of simulations with three values of gravity ($g_0=50$, $100$, $150$ m s$^{-2}$) and changing 
magnetic field amplitude, $B_0$; for each value of $g_0$ the magnetic field varies in the range $1$ to $10^{-2}$ T. 
The parameters are the same as in Table~\ref{tab.1}. The growth rates marked with $^*$ are compatible with zero, 
as the relative simulations have never reached the linear growth phase after about 15 $t_{\rm A}$ and 
they show a noisy behaviour. We decided to include these values anyways for completeness.
We report the value of $B_p/B_t $ only for the simulations that
reached the end of the linear growth.
\label{tab.2}}
\end{table*}

\begin{figure*}
\begin{center}
\includegraphics[width=1.8\columnwidth]{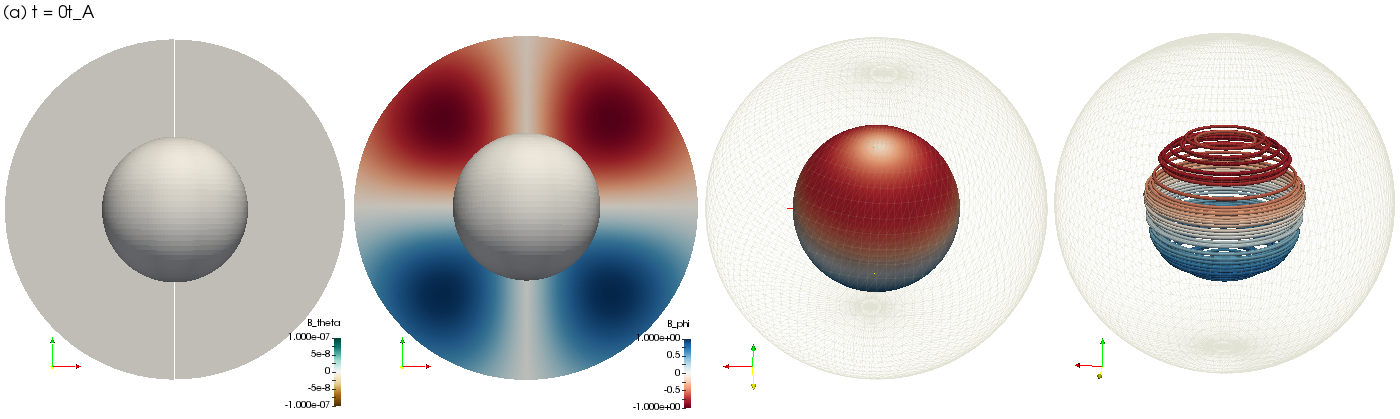}\\
\includegraphics[width=1.8\columnwidth]{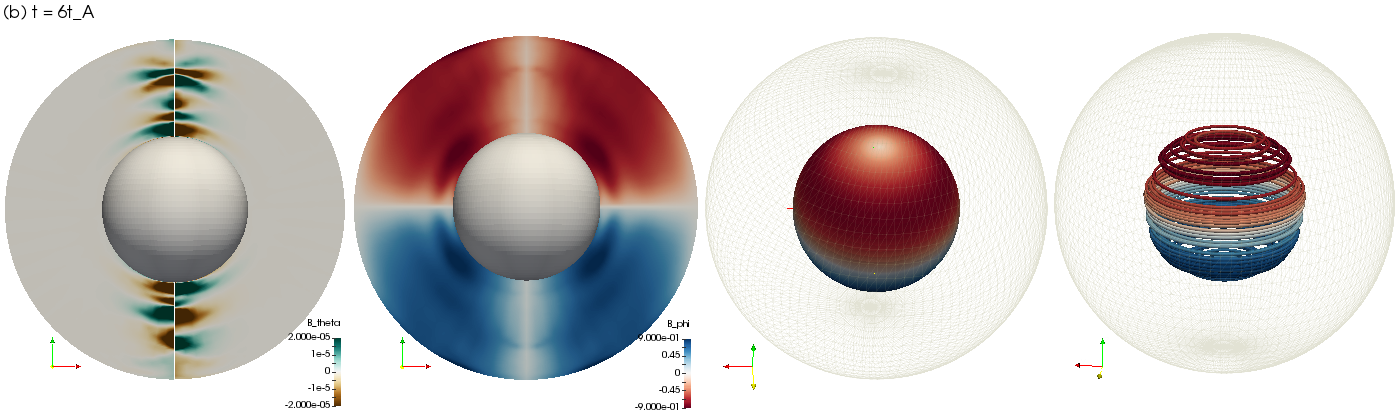}\\
\includegraphics[width=1.8\columnwidth]{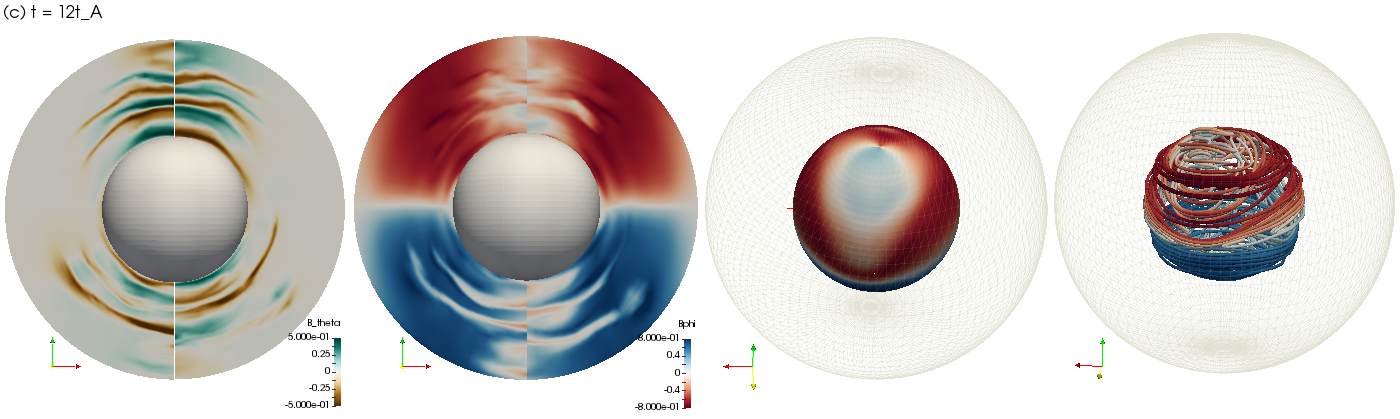}\\
\includegraphics[width=1.8\columnwidth]{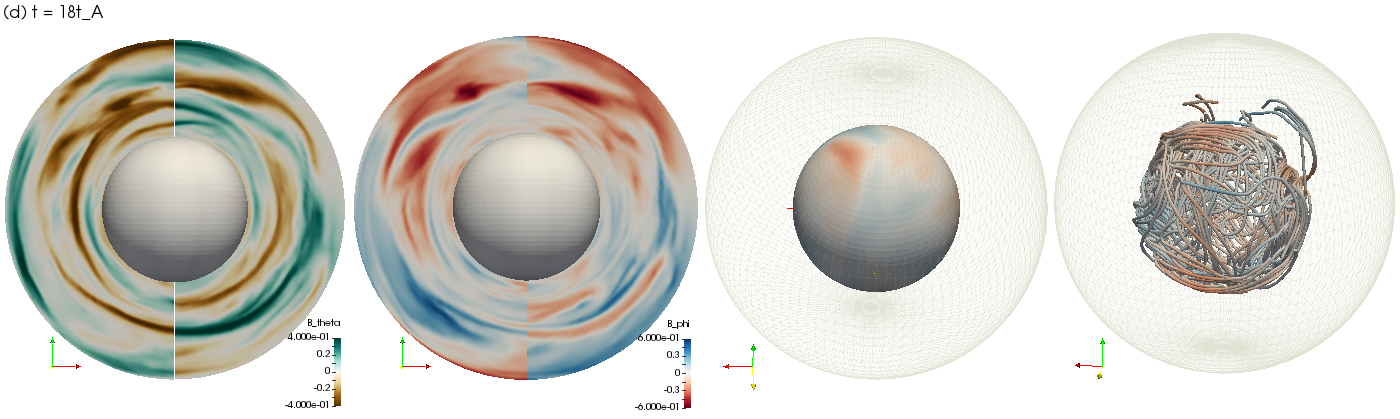}\\
\caption{Magnetic field evolution for the simulation TiA01. The first and second
columns show contourplots of the latitudinal, $B_{\theta}$, and longitudinal, $B_{\varphi}$,
components of the magnetic field in the longitudinal, $r - \theta$, plane for $\varphi = 0$ 
and $\pi$. 
The third column shows $B_{\varphi}$ on the $\varphi-\theta$ 
surface at $r=0.45 R_{\rm s}$. 
The fourth shows the magnetic field 
lines at $r=0.45 R_{\rm s}$: the red (blue) color represents a
counterclockwise (clockwise) longitudinal field.
Different rows correspond to different evolution times
in Alfvén travel times:  $t=0t_{\rm A},~ 6t_{\rm A},~ 12t_{\rm A},~ 18t_{\rm A}$. 
}
\label{f.fe01}
\end{center}
\end{figure*}

\begin{figure*}
\begin{center}
\includegraphics[width=1.8\columnwidth]{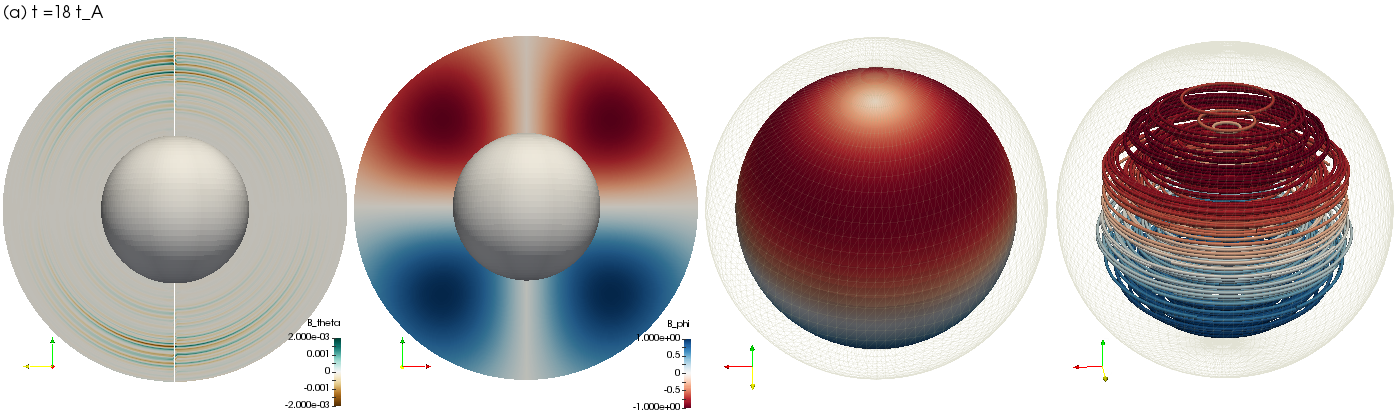}\\
\includegraphics[width=1.8\columnwidth]{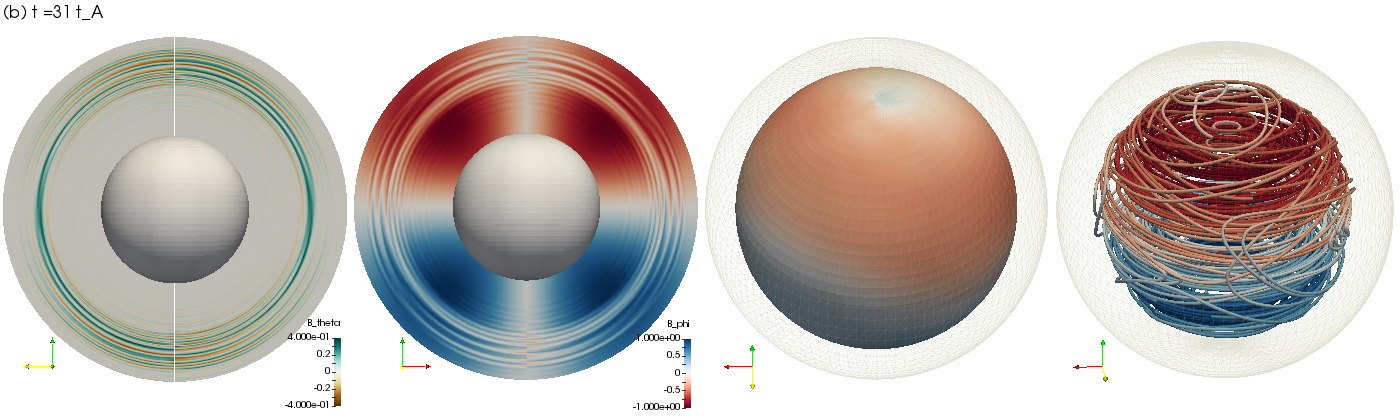}\\
\includegraphics[width=1.8\columnwidth]{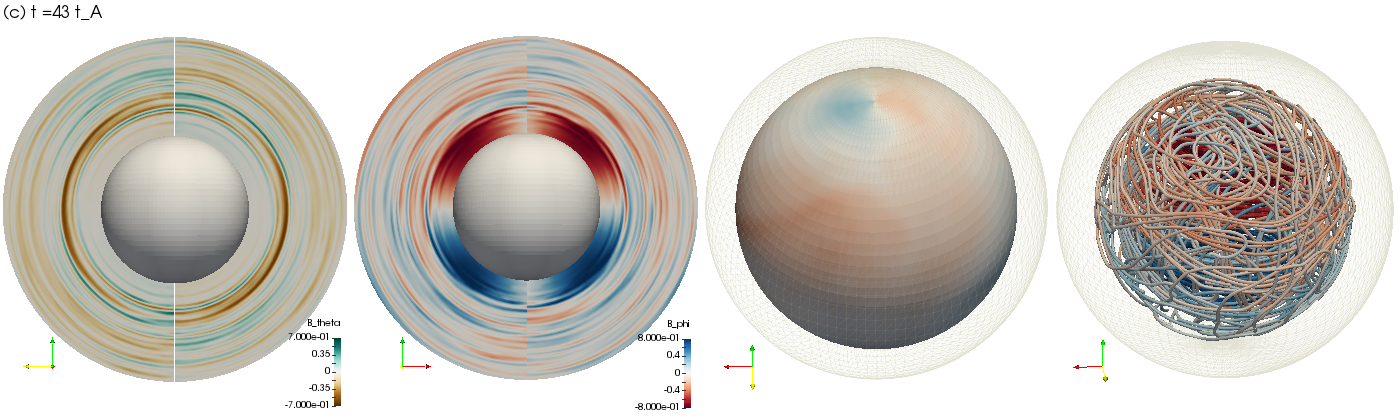}\\
\caption{Same as Fig.~\ref{f.fe01}, but for simulation TiA05 at $t= 18t_{\rm A},~ 31t_{\rm A},~ 43t_{\rm A}$.
In this case the field
configurations in third and fourth columns are shown at $r=0.85 R_{\rm s}$.}
\label{f.fe04}
\end{center}
\end{figure*}

\section{Global numerical simulations}\label{Sec:3D}
To overcome the limitations of the earlier analytical considerations
we perform global numerical simulations with the EULAG-MHD code.
We study the non-rotating case by solving the anelastic MHD equations in the following form:
\begin{equation}
{\bm \nabla}\cdot(\rho_{\rm ad} \bm u)=0, \label{equ:cont}
\end{equation}
\begin{equation}
        \frac{D \bm u}{Dt} =  
    -{\bm \nabla}\left(\frac{p'}{\rho_{\rm ad}}\right) + {\bm g}\frac{\Theta'}
     {\Theta_{\rm ad}} + \frac{1}{\mu_0 \rho_{\rm ad}}({\bm B} \cdot \nabla) {\bm B} \;, 
\label{equ:mom}
\end{equation}
\begin{equation}
        \frac{D \Theta'}{Dt} = -{\bm u}\cdot {\bm \nabla}\Theta_{\rm amb} -\frac{\Theta'}{\tau}\;, \label{equ:en}
\end{equation}
\begin{equation}
 \frac{D {\bm B}}{Dt} = ({\bm B}\cdot \nabla) {\bm u} - {\bm B}(\nabla \cdot {\bm u})  \;,
 \label{equ:in}
\end{equation}
\noindent
where $D/Dt = \pd/\pd t + \bm{u} \cdot {\bm \nabla}$ is the total                                                                  
time derivative, ${\bm u}$ is the velocity field,
$p'$ is a pressure perturbation variable that accounts for both the gas
and magnetic pressure, and ${\bm B}$ is the magnetic field. 
The energy equation (\ref{equ:en}) is written in terms of perturbations of the potential
temperature, $\Theta'$,
with respect to the ambient state, $\Theta_{\rm amb}$.  
The latter is chosen to be 
roughly isothermal 
\citep[see][for comprehensive discussions about this formulation of the energy 
equation]{GSKM13b,Cossette+17}.
The $\rho_{\rm ad}$  and $\Theta_{\rm ad}$ are the density and potential 
temperature of the reference isentropic state
(i.e., $\Theta_{\rm ad}={\rm const}$)
in hydrostatic equilibrium: the potential temperature, $\Theta'$, is related to the specific
entropy by 
$ds=c_p  d \ln\Theta'$; $g = \frac{g_0}{(r/r_b)^2}\bm{\hat{e}}_r$  
is the gravity acceleration, with $g_0$ its value at the bottom of the domain where $r=r_b$;
and $\mu_0$ is the magnetic permeability of the vacuum;
the last term in equation~\ref{equ:en} is a Newtonian cooling that relaxes
$\Theta'$ in a time scale $\tau=1.296\times 10^7$ s.
For the simulations presented here 
this time scale is shorter than the average Alfvén travel time
$t_{\rm A}= 2\pi/\omega_{A}$ (see Eqs. \ref{eq.ndq} and \ref{eq.om_A}).

The Newtonian cooling acts in the simulations as scale independent thermal 
diffusion that substitutes the thermal and radiative diffusion expected to 
exist in stellar interiors.  The value that we use is a compromise between having a
fast cooling or a slow thermal diffusion given only by the numerical resolution,
therefore allowing large values of $\Theta'$.  
We verified the effects of this term by running auxiliary simulations (not shown)
with different values 
of $\tau$.  In cases where the Newtonian cooling is 100 times shorter,
i.e.,  $\tau=1.296\times 10^5$ s, 
the instability is partially suppressed.  In simulations where the 
Newtonian cooling term is removed (i.e., $\tau \rightarrow \infty$)
we observe an overall behaviour similar to
the simulations with our fiducial value of $\tau=1.296\times 10^7$ s. Nevertheless,
the average values of the growth rate of the instability are somewhat smaller.
This is consistent with
the findings of the linear theory by \cite{bu12apj}, and demonstrate that the
thermal diffusivity enhances instability. 

We consider a spherical shell with $0\le \varphi \le 2\pi$,
$0\le \theta \le \pi$ and a radial extent from $r_b=0.4\Rs$ to $r_t=0.96\Rs$.
The boundary conditions are defined as follows: for the velocity
field we use impermeable, stress-free conditions at the top and bottom surfaces
of the shell; for the magnetic field we consider a perfect conductor at both
boundaries. Finally, for the thermal boundary condition we consider zero
radial flux of potential temperature.
The discrete mesh in most of the simulations has $126\times42\times72$ grid
points in the $\varphi$, $\theta$ and $r$ directions, respectively. The constant time-step
of the simulations is $\Delta t = 1800 $ s. For the high resolution simulations we
double the number of grid points in each direction and decrease the time-step to
$\Delta t = 450$ s.
The reference and ambient states are computed by solving the hydrostatic
equilibrium equations for a polytropic atmosphere,
\ba
\frac{\partial T_{\rm i}}{\partial r} &=& -\frac{g}{R_g (m_{\rm i}+1)}\;, 
\label{equ:equilT}
\\
\frac{\partial \rho_{\rm i}}{\partial r} &=& = -\frac{\rho_{\rm i}}{T_{\rm i}}
     \left(\frac{g}{R_g} - \frac{\partial T_{\rm i}}{\partial r} \right)\;,
     \label{equ:equilrho}
\ea
where the index ${\rm i}$ stands either for ${\rm ``ad"}$  or ${\rm ``amb"}$, 
$R_g=13732$ is the gas constant. Density and temperature are related to
the gas pressure through the equation of state for a perfect gas, 
$p_{\rm i} = R_g \rho_{\rm i} T_{\rm i}$.
The bottom boundary values used to integrate  (\ref{equ:equilT}) and (\ref{equ:equilrho}) 
are $T_b = 3.5 \times 10^6$ K 
and $\rho_b = 37~{\rm kg}~{\rm m}^{-3}$, respectively.
Different values of $g_0$ allow to obtain different 
degrees of 
stratification.
Finally, adiabatic and roughly isothermal 
atmospheres are obtained with the polytropic indexes $m_{\rm ad} = 1.5$ and 
$m_{\rm amb} = 10^3$, respectively. 

The simulations start with a purely toroidal magnetic field, 
\ba
B_{r0}      = 0~, \;
B_{\theta 0}= 0~, \; 
B_{\varphi 0}  = B_0 f(r) \sin 2 \theta \;,
\label{eq.b0a}
\ea
with
\be
f(r) = \exp\left(-(r - r_0)^2/d^2 \right),
\label{eq.b0b}
\ee
where $r_0 = 0.68 \Rs$, 
$d = 0.5 \Rs$, 
and $B_0$ is the maximum amplitude of the
initial magnetic field which is a free parameter in the simulations (see Table \ref{tab.1} and \ref{tab.1}). 

To directly compare the 3D simulations with the linear analysis we consider 
the non-dimensional quantity $\delta^2$, defined in Eq.~(\ref{eq.ndq}). 
However, in the global models  we have a gravity profile depending on 
radius, and an initial magnetic field depending on $r$ and $\theta$;
see Fig.~\ref{f.delta}. Thus, we consider
$\mean{\delta}^2 = \mean{\omega}_{BV}^2/ \mean{\omega}_A^2$, with
\be
\mean{\omega}_{BV}^2 = \brac{ \frac{g}{\Theta_{\rm amb}} \frac{\partial \Theta_{\rm amb}}{\partial r}}_r   , \; \;
\mean{\omega}_{A}^2  = \brac{ \frac{B_{\varphi 0}^2}{\mu_0 \rho_{\rm amb} d^2} }_{r,\theta}.
\label{eq.om_A}
\ee
Here, the angular brackets represent averages in the radial direction for the \bvf and
in radius and latitude (over one hemisphere) for the Alfvén frequency.
The values of  $\mean{\omega}_{BV}$, $\mean{\omega}_{A}$ and $\mean{\delta}$ 
are presented in Table~\ref{tab.1}.

By construction our initial state is Tayler unstable. Therefore, it is expected that the 
instability develops after a few characteristic Alfvén travel times $t_{\rm A}$. 
As we will see, this occurs quickly for the 
models with strongest initial magnetic fields. On the other hand, the simulations with 
weaker fields ($B_0 \le 0.5$ T) reach magnetohydrostatic equilibrium after about one $t_A$.
Therefore, to excite the instability we impose a 
white noise perturbation 
with amplitude of $10^{-3}$ m s$^{-1}$
and continue the simulations for, at least, 
$20 t_{\rm A}$.

\begin{figure}
\begin{center}
\includegraphics[width=0.98\columnwidth]{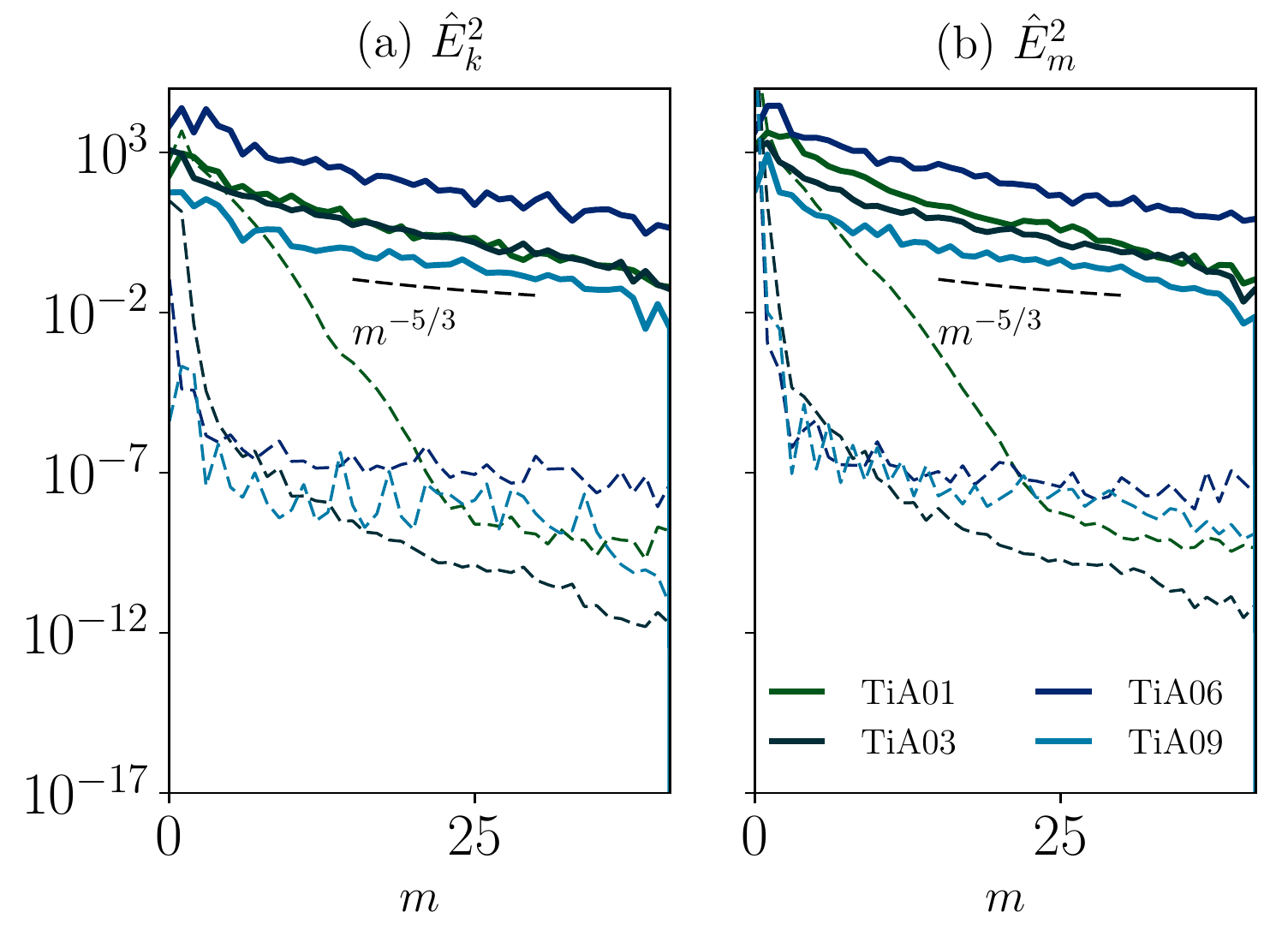}
\caption{  Kinetic (a) and  magnetic (b) energy density spectra as a function of the 
longitudinal
wave number $m$, measured during the linear (dashed lines) and the dissipative (solid) phases,  
of some characteristic simulations. The black dashed lines correspond the $m^{-5/3}$ scaling
law.}
\label{f.mmodes}
\end{center}
\end{figure}

\section{Results}
\label{Sec:res}

In the first set of simulations we start with a strong initial magnetic
field, $B_0 = 1$ T, and vary the gravity at the bottom of the domain, $g_0$, from
$0.01$, that means that the density is roughly constant in radius, 
to $500$ m s$^{-2}$, corresponding to a 
density contrast $\sim 5.5$. 
In the weak gravity cases, 
$g_0=0.01$ and $1$ m s$^{-2}$, we observe the instability to develop fast,
on a time scale of the order of one $t_{\rm A}$.
 Figure~\ref{f.fe01} depicts the temporal evolution
of the latitudinal, $B_{\theta}$, and longitudinal,  $B_{\varphi}$, magnetic field
components for the simulation TiA01, with $g_0=1$ m s$^{-2}$. 
The first and second columns show, respectively, contours of 
$B_{\theta}$ and $B_{\varphi}$ in the meridional plane, $r-\theta$, for $\varphi=0$ 
and $\pi$. The third column shows $B_{\varphi}$ in the horizontal, $\varphi-\theta$, 
plane at $r=0.45R_{\rm s}$.  
Red (negative) and blue (positive) contours correspond 
to counterclockwise and clockwise toroidal fields, respectively.  The right column 
shows the  magnetic field lines at $r=0.45R_{\rm s}$ colored according to the 
amplitude of $B_{\varphi}$. The upper row (a) corresponds to the initial configuration; 
Eq.~(\ref{eq.b0a}). The second 
row (b) corresponds to the linear phase of the instability at $t=6t_{\rm A}$. 
The development of $B_{\theta}$ with a $m=1$ symmetry starting from
the axial cylinder is evident. The decay of $B_{\varphi}$ occurs in this phase predominantly 
in the $m=0$ mode which still has larger energy. Thus, the magnetic field lines
do not show any significant change with respect to the initial state.  
The third row (c) corresponds to the end of the linear phase, 
$t=12t_{\rm A}$. At this stage the non-axisymmetric mode, $m=1$, has a larger
energy which is evident in $B_{\theta}$ as well as in $B_{\varphi}$. The instability 
seems to be occurring in different radial layers of magnetic field and 
propagating from the 
vertical axis of the sphere, $\varphi = \theta = 0$.
towards equatorial latitudes. The rightmost columns show 
the field morphology in the innermost layers. 
The bottom row (d) corresponds to the beginning of the dissipative phase. It is clear
in the panels of this row that although high order modes develop, the mode $m=1$ still
prevails at some radial levels.

Figure~\ref{f.fe04} depicts the development of the Tayler instability 
for simulation TiA05, with $g_0 = 50$ m s$^{-1}$. In this figure, however, the third 
and fourth columns show the magnetic field at $r=0.85 R_{\rm s}$. Since the initial 
configuration  is the same for all cases, we present only the magnetic field
after (a) $\sim 18 t_{\rm A}$, corresponding to the linear phase, (b) 
$\sim 31 t_{\rm A}$, corresponding to the saturated phase and (c) $\sim 43 t_{\rm A}$ when 
the field is in the diffusive decaying stage.  Similar to the case 
TiA01, the instability starts close to the 
axis 
and propagates towards the equator. 
Nevertheless, it occurs in a larger number of radial layers with different growth
rates. The panels in the bottom row indicate that the instability is fully developed at 
the upper radial levels but at the bottom of the domain the field conserves some
coherence.  The magnetic field lines, fourth column, show a fully mixed magnetic field
at $r=0.85 R_{\rm s}$ but suggest a better organized field in deeper layers.  

In Fig.~\ref{f.mmodes} we present the energy density spectra of the velocity and 
magnetic fields at the linear (dashed lines) and the dissipative (solid lines)  phases,
for some characteristic simulations  \citep[the decomposition in spherical harmonics
was performed with the optimized library SHTns][]{shtns} .  
We observe that in the simulation TiA01, characterized by
small $\mean{\omega_{\rm BV}}$,
during the linear phase most of energy is stored in about 20 
longitudinal modes.
As $\mean{\omega_{\rm BV}}$ increases, the number of  longitudinal
modes decreases.
On the other hand, when the system reaches the dissipative phase, we 
observe the occurrence of fully developed 3D turbulence independently of the stratification:
this is illustrated by the 
behavior of the kinetic and magnetic energy power spectra, which exhibit
a $m^{-5/3}$ power law decay.  

\begin{figure}
\begin{center}
\includegraphics[width=0.98\columnwidth]{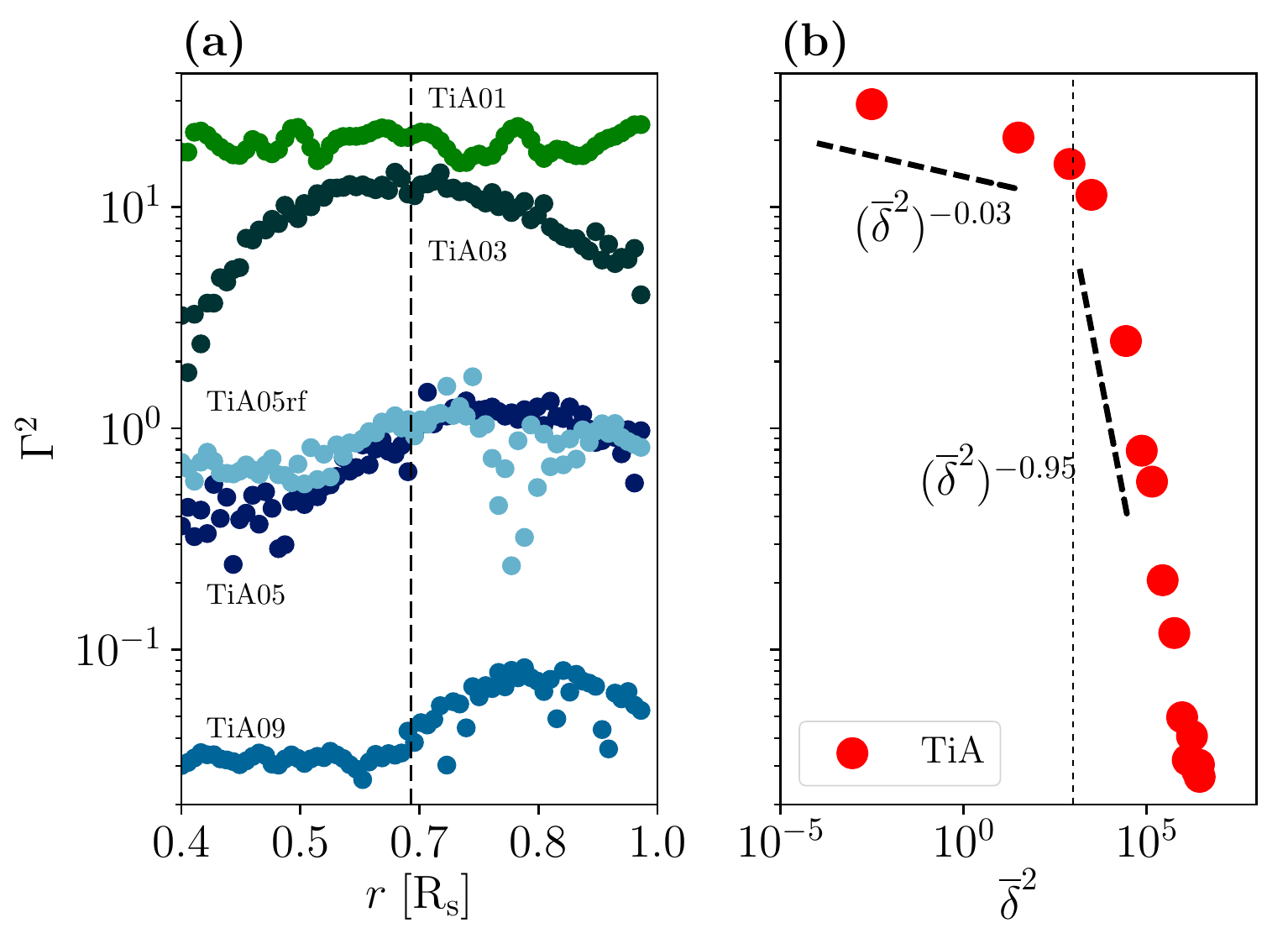}
\caption{(a) Growth rate squared, $\Gamma^2$, of the Tayler instability as
a function of radius for some characteristic simulations with different 
$g_0$. The vertical dashed line corresponds to the radius where the initial magnetic field
is maximum. 
(b) $\Gamma^2$ as a function of $\delta^2$ for the set of simulations TiA. The growth rate
corresponds to $r=0.68 R_{\rm s}$. Here the vertical dashed line shows
approximatively the separation between the two regimes
$\Gamma^2 \propto \delta^{-0.03}$
and
$\Gamma^2 \propto \delta^{-0.95}$.
}
\label{f.grk}
\end{center}
\end{figure}

\begin{figure}
\begin{center}
\includegraphics[width=0.98\columnwidth]{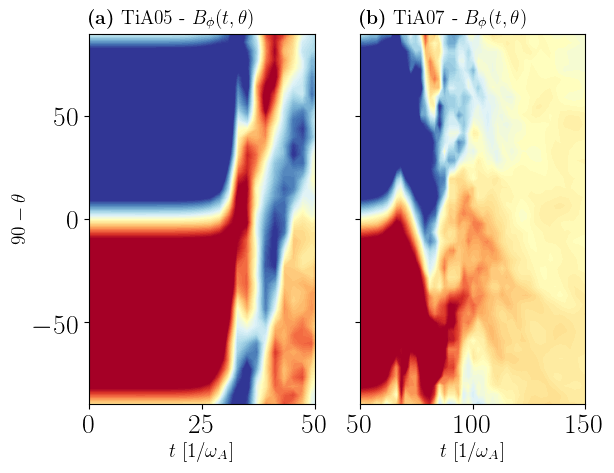}
\caption{Time-latitude evolution of the toroidal magnetic field at $\varphi = 90^{\circ}$
and $r=0.72 R_s$ for the simulations (a) TiA05 and (b) TiA07.
}
\label{f.but}
\end{center}
\end{figure}

\begin{figure*}
\begin{center}
\includegraphics[width=1.9\columnwidth]{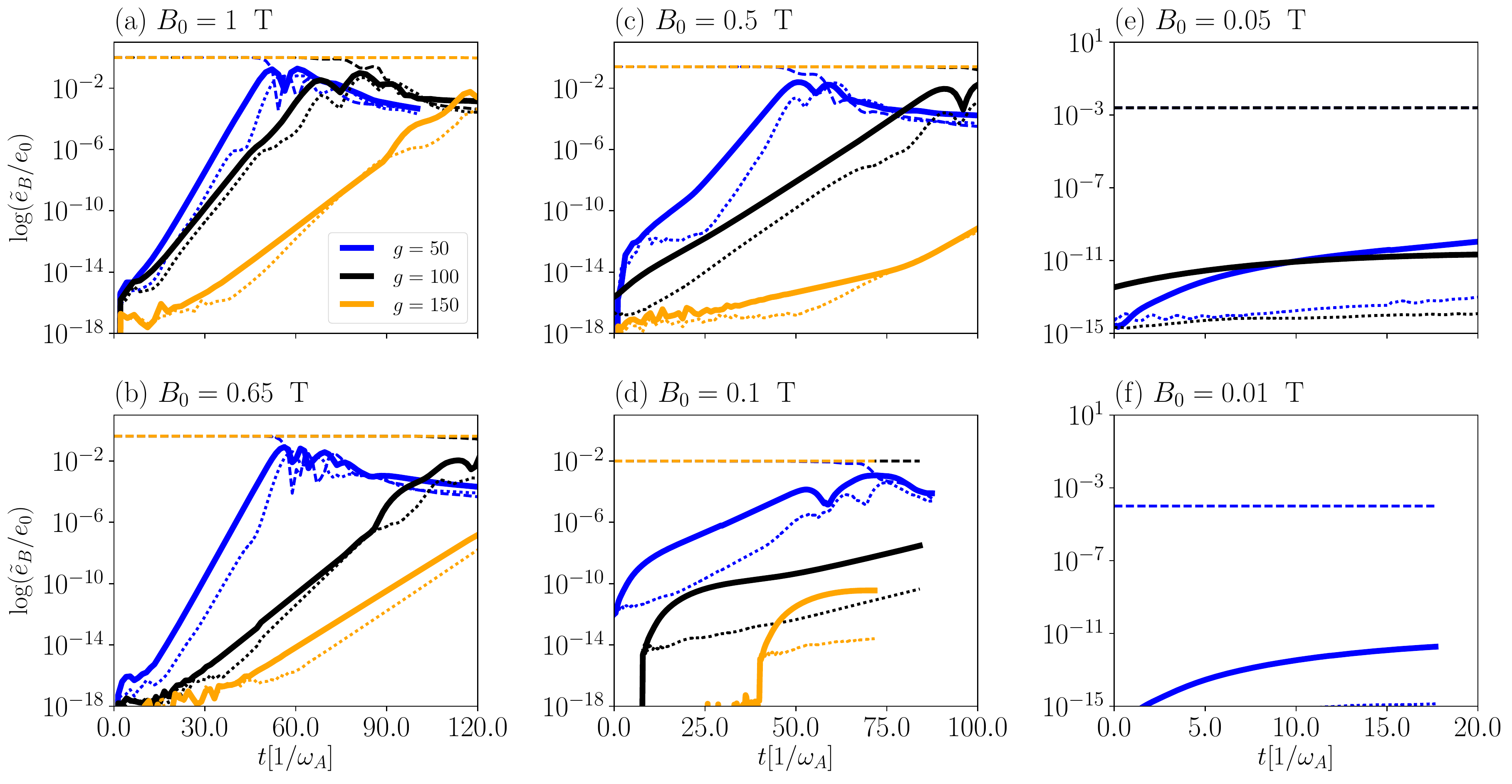}
\caption{Time evolution of the simulations in Table \ref{tab.2}. Different colors 
correspond to different values of $g_0$ as shown in the legend. The dashed, solid and
dotted lines correspond to the modes $m=0, 1$ and  $2$, respectively.}
\label{f.multi}
\end{center}
\end{figure*}

\begin{figure}
\begin{center}
\includegraphics[width=0.98\columnwidth]{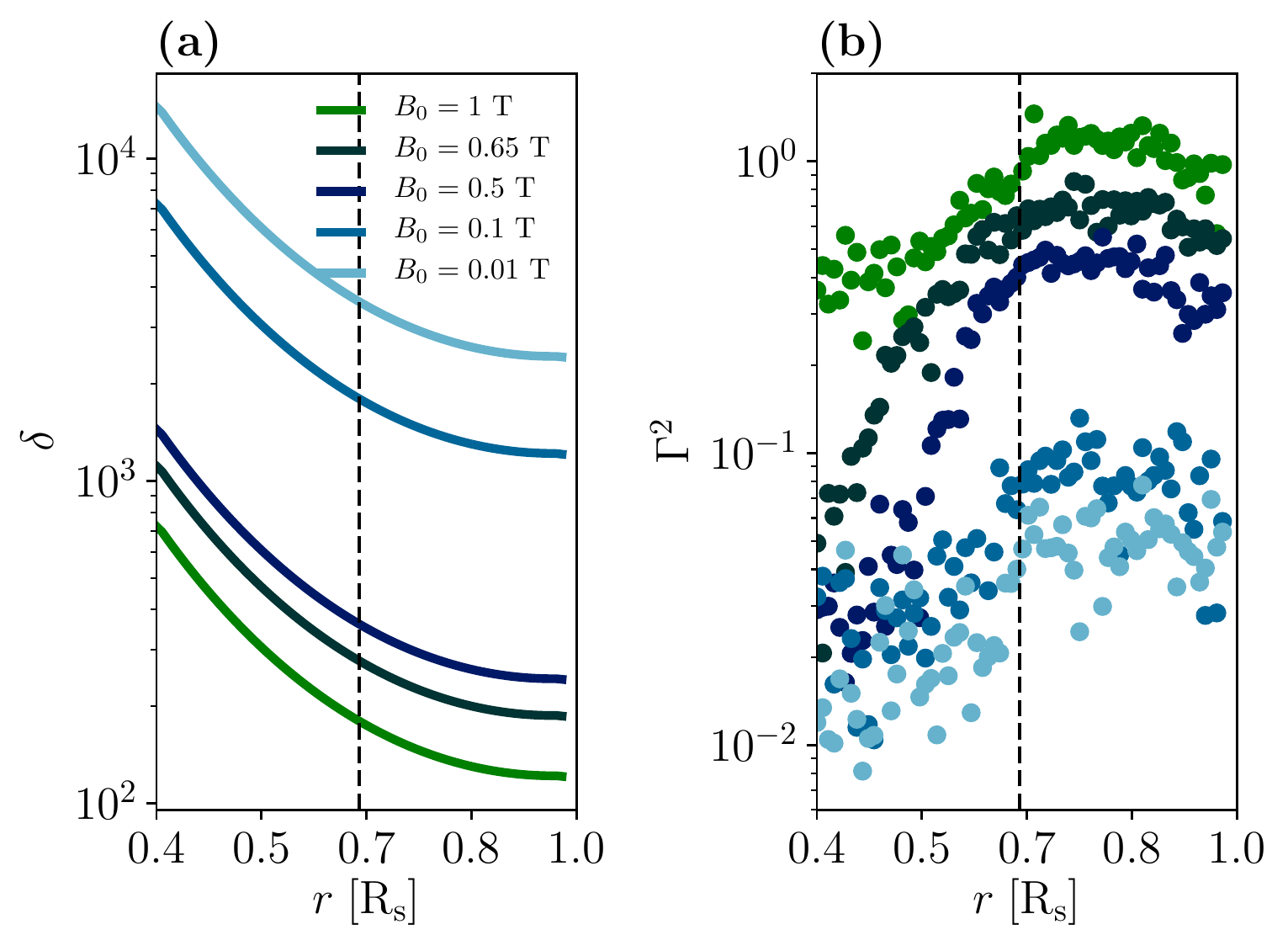}
\caption{(a) Radial profile of $\delta$ for simulations with fixed
$g_0 = 50$ m s$^{-2}$, and magnetic field with maximum amplitude, $B_0$ between
$1$ and $0.01$ T.  (b) Growth rate squared, $\Gamma^2$, as
a function of radius for these simulations. 
The vertical dashed lines corresponds to the radius where the initial magnetic field
is maximum. }
\label{f.grk50}
\end{center}
\end{figure}

\begin{figure}
\begin{center}
\includegraphics[width=0.9\columnwidth]{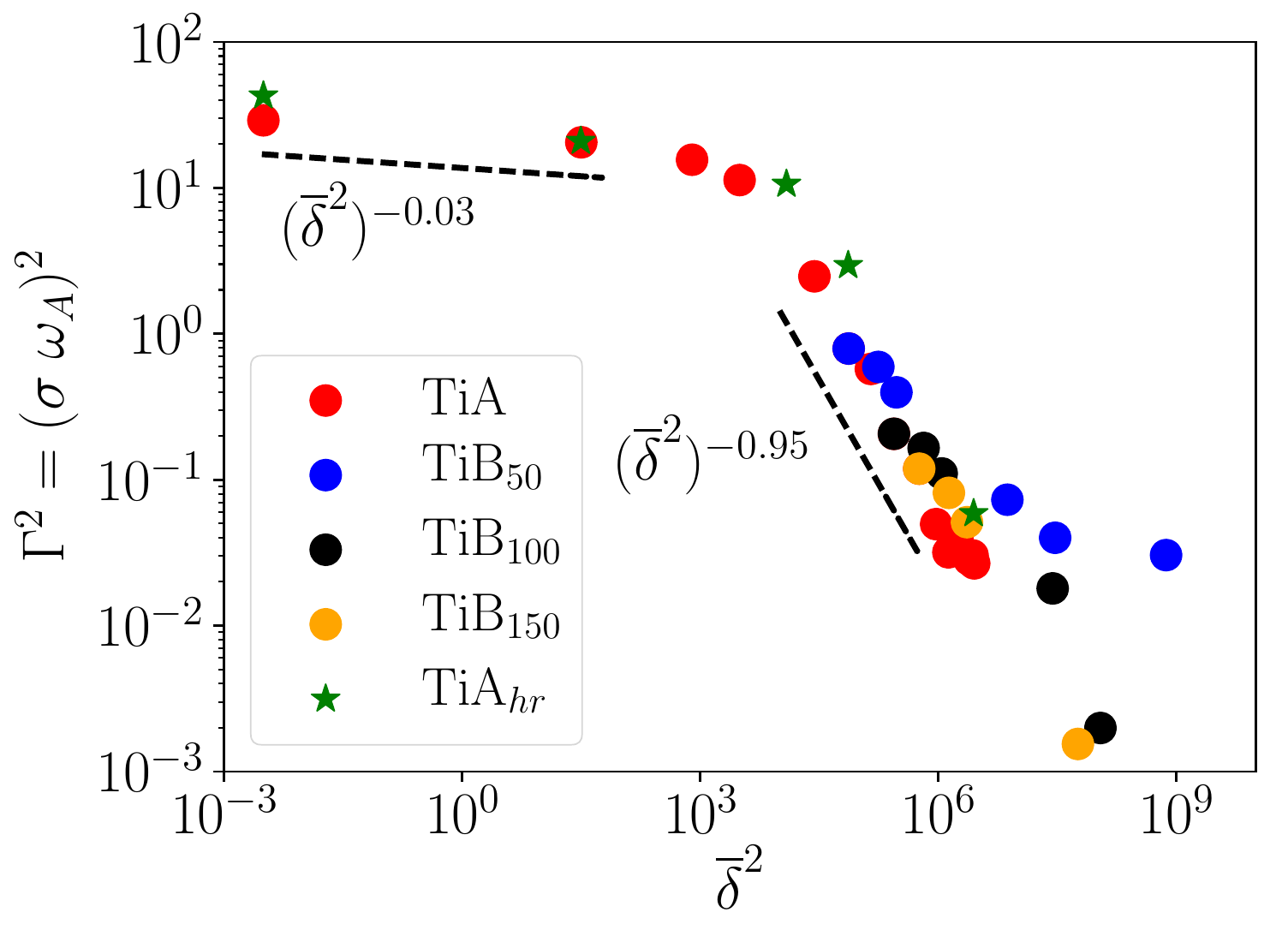}
\caption{Dimensionless growth rate squared for the mode $m=1$, as a function of 
$\mean{\delta}^2$ for the models in Tables~\ref{tab.1} and \ref{tab.2}.}
\label{f.lvd}
\end{center}
\end{figure}

Since the development of the Tayler instability occurs at different time scales
for different depths, we compute the growth rate, $\Gamma$, 
of the instability
at each radial point. 
$\Gamma$ is estimated 
from the time evolution of the mode $m=1$ of the magnetic 
field energy density. In Fig.~\ref{f.grk}(a) we present the growth rate of the Tayler instability,
in Alfvén travel times $t_{\rm A}$,  as a function of radius for some characteristics 
simulations with different values of $g_0$.
We notice that the growth rate of the instability
is roughly independent of radius for small $g_0$, i.e.,  small $\mean{\delta}$. 
Yet, when 
$\mean{\omega_{\rm BV}}$
increases, the growth rate is smaller at the bottom of the domain 
where $\delta$ has larger values. It reaches a maximum at a 
radius that seems to increase with $g_0$, and then decreases again 
near the upper boundary.  For the cases with larger $g_0$ (simulations
TiA09-TiA14), the growth rate is constant in the lower half of the domain; 
e.g., see the points corresponding to simulation TiA09 in Fig.~\ref{f.grk}(a)).
To study the role of the magnetic boundary condition on the growth rate we performed
one simulation, TiA05rf, 
with pseudo-vacuum boundary condition at the upper
boundary.  
Even though the radial profile of $\Gamma$ is slightly different, 
on average the 
growth rate agrees with that of the simulation with perfect conductor boundary condition
(see light blue points in Fig.~\ref{f.grk}(a)
as well as Fig.~\ref{f.lvd}).

Figure~\ref{f.grk}(b) shows $\Gamma^2$ at $r=0.68R_{\rm s}$ (where the initial toroidal 
magnetic field has a maximum) as a function of $\mean{\delta}$ for all simulations 
with $B_0 = 1$ T.  The growth rate  follows two different behaviors.
For $\delta \lesssim 50$ (see the vertical dotted line) the growth rate is large and decays slowly 
following the power law $(\delta^2)^{-0.03}$. For $\delta \gtrsim 50$ the decay of the growth rate
is fast and follows the power law $\Gamma^2 \propto (\delta^2)^{-0.95}$.
In this region $\Gamma^2$ changes by a factor of $10^2$.  For larger values of $g_0$;  
($\mean{\delta}^2 \gtrsim 10^6$), the growth rate remains approximately constant. This is 
expected since increasing $g_0$ above 200 m~s$^{2}$ 
results in adjacent BV frequency profiles as can be seen in Fig.~\ref{f.delta}. 
The trend  $\Gamma^2 \propto (\delta^2)^{-0.95}$ depicted in Figs.~\ref{f.grk}(b) and \ref{f.lvd} is consistent 
with the findings of \cite{bu12apj}, who used eq.(\ref{stab}) and obtained a
stabilizing effect of gravity with a power law $\Gamma^2 \propto (\delta^2)^{-1}$.
Also, $\Gamma$ can attain values smaller than 1, in agreement with
findings of \cite{GTZ2019}

In Table~\ref{tab.1} we also report a measurement of 
the ratio of mean poloidal and toroidal fields, calculated 
over the entire domain at the end of the linear growth of the instability. 
Whilst we could not identify an explicit dependence of this ratio on the initial values 
of the parameters $g_0$ and $B_0$, it is clear that a poloidal component of the field
develops, and it keeps growing during the nonlinear phase.

\subsection{The role of initial magnetic field strength}

To further increase the value of $\mean{\delta}$ we run three sets
of simulations where $g_0$ is either $50$, or $100$ or $150$ m s$^{-2}$.
For each of them, we explore the role of the initial magnetic field strength, by varying its
maximum amplitude, $B_0$, from $1$ down to $0.01$ T.  

The results of these simulations are presented in Table~\ref{tab.2}, and 
the time evolution for the these sets is depicted in Fig.~\ref{f.multi}(a)-(f).
Each panel shows the results for different field amplitudes, with the blue, 
black and yellow lines corresponding to $g_0=50$, $100$ and $150$ m s$^{-2}$,
respectively. The dashed, solid and dotted lines correspond to the 
longitudinal modes, $m=0$, $1$ and $2$, respectively. 
The results confirm that the behavior described above for $B_0 = 1$ T
holds also for smaller values of the initial field,
i.e., the stronger the gravity force, the smaller the
instability growth rate. As mentioned above, for large values of $B_0$ the instability 
starts developing after 1-3 $t_{\rm A}$, without the need of perturbing 
the system.  For $B_0 \leq 0.1$ T the initial magnetic field remains 
stationary while the velocity field and the potential temperature adapt to
magneto-hydrostatic equilibrium. We run these initial states for $\sim10 \; t_A$.
The instability develops after perturbing the system with white noise in the 
potential temperature perturbations.
For some of the simulations, instead, the instability doesn't develop even after perturbing
the system, illustrating how the system is stabilized by the combination of strong gravity and weak fields.
Figure~\ref{f.multi}(a)-(d) show that in most of the simulations 
the end of the linear growing phase is characterized by oscillations. These patterns
resemble the results by \cite{weber2015}, who found helicity oscillations 
in simulations of the Tayler instability in cylindrical coordinates. In our case
we notice that they correspond to waves of magnetic field that travel from one
pole to the other. Figure~\ref{f.but}(a) and (b) show the time evolution of the
toroidal magnetic field of simulations TiA05 and TiA07 to illustrate this pattern. 
The toroidal field is sampled at a longitude
of $90^{\circ}$ and a radius of $r=0.76 R_s$.
The migration direction might change for different choices of longitude and is 
more evident at a radius closer to the maximum of the initial toroidal field. 
The oscillations seem to have longest period for simulations with small 
$\mean{\delta}$ and are not clearly defined in simulations with 
$\mean{\delta} \gtrsim 10^3$. This pattern rapidly disappears
once the field enters in the dissipative phase.

In Fig.~\ref{f.grk50} we compare the radial profiles of (a) $\delta$,
and (b) $\Gamma^2$ as function of radius in simulations with $g_0 = 50$ m s$^{-2}$
and different values of $B_0$.
The figure indicates that for all the cases the instability growth rate is 
smaller at the bottom (where $\delta$ is larger) and larger and the top
(small $\delta$). Also, the maximum of the growth rate 
appears roughly at the same radius, $r \sim 0.75 R_{\rm s}$, in all simulations. 
Nevertheless, while for the smaller values of $\delta$ (i.e., for $B_0 > 0.5$ T) 
the profiles of $\Gamma^2$ have a clear radial trend, for higher
$\delta$ (i.e., $B_0 \leq 0.1$ T) the growth rate shows a significant dispersion. 
For this reason, the growth rate presented in Table~\ref{tab.2} corresponds
to the radial average of $\Gamma$ between $0.66 r/R_{\rm s}$ and $0.7 r/R_{\rm s}$.

The combination of strong gravity and weak magnetic field
allows to reach $\mean{\delta^2} \sim 10^9$ (Table~\ref{tab.2}),
These results are presented in Fig.~\ref{f.lvd} where the red points
are the same presented in Fig.~\ref{f.lvd}, and the blue, black and yellow 
points correspond to $g_0=50$, $100$ and $150$ m s$^{-2}$, respectively.
The results suggest that for each set of simulations in Table~\ref{tab.2} there is 
a power law decay. Nevertheless, for the large values of $g_0$ considered (i.e., 
$100$ and $150$ m s$^{-2}$), the exponent seems to be similar. 
The figure clearly evidences that gravity inhibits the instability of the magnetic
field by changing $\Gamma^2$ by several orders of magnitude.

The combination of strong gravity ($g_0=100$, $150$ m s$^{-2}$) and 
weak fields ($B_0 \leq 0.05$ T) results in growth rates almost negligible.
This is a consequence of the stabilizing effect of gravity.   We notice, however,  
that simulations with small $B_0$, are characterized by longer Alfvén travel times.
Thus, these simulations require a much longer computational time which at the 
moment is prohibitive. So these simulations have not reached saturation, and 
the possibility that the systems will become unstable on scales of the order of
$t\sim100 t_A$ cannot be {\it apriori} excluded. 
Nonetheless, there is a clear trend showing that weaker magnetic 
fields are stable on longer timescales. 

\subsection{Effects of resolution}
We also explore the role of resolution in our simulations by doubling 
number of cells in all directions, i.e., considering $252\times84\times144$
grid points in $\varphi$, $\theta$ and $r$, respectively. 
We observe that the values of $\Gamma$ for each $\mean{\delta}$ 
are slightly larger than their low-resolution counterparts
(see simulations TiAhr00-TiAhr14 in Table~\ref{tab.1}).  
Nonetheless, it is interesting to 
notice that in the simulations with higher resolution the trend of $\Gamma$ as 
a function $\mean{\delta}$ seems to be the same as for the lower resolution case. 
The two regimes described above are discernible in Fig.~\ref{f.lvd}
(see the green stars). This confirms that our numerical approach is able to capture 
the stabilizing effect of gravity.

\begin{figure}
\begin{center}
\includegraphics[width=0.49\textwidth]{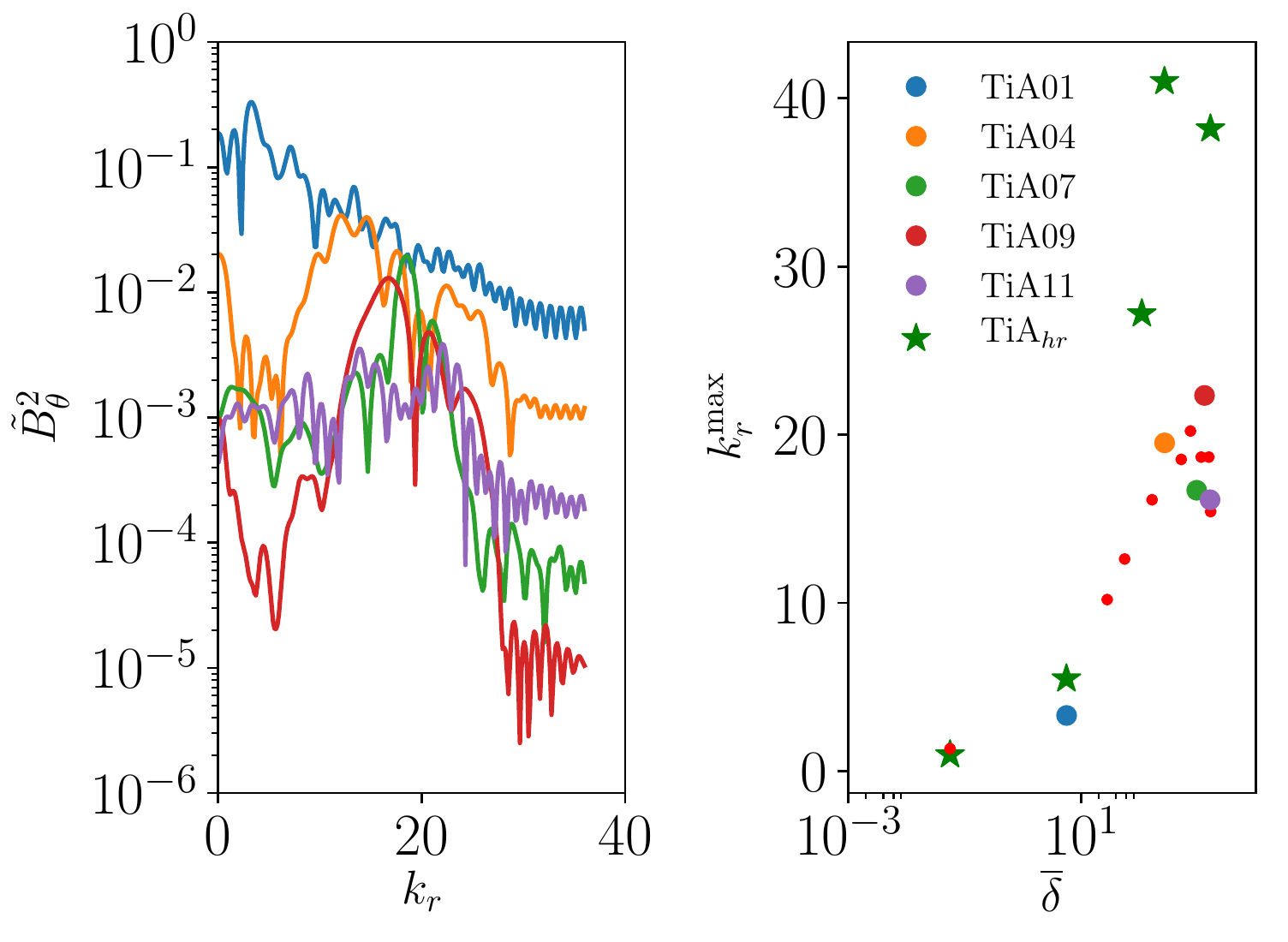}
\caption{Left: Fourier energy spectra in the radial direction as a function of the
vertical wave number, $k_r$,  for some characteristic simulations. Right: Maximum 
vertical wave number as function of $\mean{\delta}$. 
The color points correspond to the simulations indicated in the legend.
The red points show the results for the all other low resolution simulations in Table~\ref{tab.1}, 
whilst the green stars correspond to the high-resolution simulations.  
}
\label{fig.rmodes}
\end{center}
\end{figure}

\subsection{Radial modes}

While studying the temporal evolution of the simulations we noticed
that increasing the value of $\mean{\delta}$ by increasing $g_0$ leads to a
larger number of radial modes (see Figs.~\ref{f.fe01} and \ref{f.fe04}).
We quantify this number by computing the spectra of kinetic and magnetic energy 
density through a Fourier analysis.  The results presented in Fig.~\ref{fig.rmodes} 
and in Tables~\ref{tab.1} and \ref{tab.2}
shows that the 
radial wave number, $k_r$, increases with delta following a
power law. However, for the $g_0 \gtrsim 100 $ the number of radial modes
oscillates around $\sim 18$.  This is expected since the values and 
profiles of $\delta$ do not significantly change for these gravity values,
similarly, the growth rates are all around the same values.  

For simulations with constant $g_0=50, 100$ or $150$, and varying initial 
field strength, $B_0$, the number of radial modes is similar, fluctuating about 18
(for clarity these simulations have not been included in Fig.~\ref{fig.rmodes}). 
On the other hand, the simulations with high-resolution develop a number
of radial modes that doubles that of the low-resolution cases at the same $\delta$ 
(see the green stars in Fig.~\ref{fig.rmodes}). 
We also notice that the number of modes changes with time,
but for each simulation we can identify a portion of the linear phase of the instability
where the number of modes stays almost constant, and it attains the same
value both for the magnetic and the velocity fields. This is the number we
report in in Tables~\ref{tab.1} and \ref{tab.2}.
These findings indicate that the Tayler instability has a radial dependence
for small values of $\mean{\omega_{\rm BV}}$ only.
When the Brunt-Väisäla frequency exceeds a value
about $10^{-4}$ Hz, the instability becomes independent of the stratification.
In other words, it becomes bidimensional in nature. 
We interpret this as a numerical effect, where
a few grid points in the radial directions are required for the initial
magnetic field to decay into a mixed, toroidal-poloidal configuration, through 
reconnection processes. 

\section{Conclusions}\label{Sec:conc}

We present global anelastic simulations of the evolution of toroidal magnetic fields 
in a stable stratified, roughly isothermal, environment. The initial field configuration is consistent
with two bands of toroidal field antisymmetric across the equator. We study the development of 
the Tayler instability and assess the role played by the Brunt-Väisäla and the Alfvén frequencies on its 
characteristics. We do this by changing the gravity, $g_0$, and the initial magnetic field strenght, $B_0$.

As expected, the fastest growing mode is $m=1$. Nevertheless, other 
longitudinal
modes also develop; 
for the lower values of $g_0$ other large scale modes develop until $m\sim20$. As $g_0$ increases 
only large scale modes grow during the 
linear phase, i.e., $m \lesssim 3$.  In the radial direction the behavior is similar, the stronger the
gravity the larger the number of modes, and the instability appears to develop roughly horizontally. 
The maximum number of radial modes, $k^{max}_r$, reaches some saturation for large $g_0$, and 
this number doubles when we double the resolution of the simulations.  We interpret this as a numerical
constrain from the inviscid numerical technique of the EULAG-MHD code for the magnetic field to evolve.
This suggests that the number of radial modes might depend on the value of the magnetic
diffusivity, $\eta$, i.e., in ideal MHD the evolution of the field in stable stratified layers 
might be bi-dimensional. 

When reaching the saturated phases of the instability the time-series of the simulations
exhibit oscillations.  We identified these as waves of magnetic field that travel from
one hemisphere to the other. After one or two periods these oscillations disappear as
the magnetic field reaches the dissipative phase.
We can speculate that these oscillations will continue and sustained dynamo action 
might appear if differential rotation is present in these layers. It should replenish 
the toroidal magnetic field, later on achieving cyclic behavior.
In the dissipative phase of the simulations presented here, fully developed MHD turbulence is
observed during this stage. Most of the energy density is in the large-scale modes, $m = 0$ 
and $1$, and an energy cascade that roughly goes as $m^{-5/3}$ is observed. 

The growth rate of the Tayler instability as a function of $\mean{\delta}$, 
averaged ratio between the BV and the Alfvén frequencies, shows two regimes. 
For $\mean{\delta} \lesssim 50$ ($g_0 \lesssim 10$), $\Gamma^2 \propto \delta^{-0.03}$. 
For larger values of $\delta$, $\Gamma^2 \propto \delta^{-0.95}$ (Fig.~\ref{f.grk}). When the value 
of $\delta$ is increased by decreasing the initial field, $B_0$, this trend seems to be 
different for low values of $g_0$, but converges to $\delta^{-0.95}$ for $g_0 =100$ and $150$
m s$^{-2}$.  In the cases of weaker initial magnetic fields we observe a clear suppression of 
the instability. The simulations do not go unstable for $B_0=0.01$ T on timescales of 
about 15 Alfvén travel times. Nonetheless we cannot exclude the instability will occur later 
on, after more than hundred $t_A$.
The transition between the two observed regimes might be connected with the
existence of a threshold for the magnetic field strength observed in Ap/Bp stars.
For a given $\omega_{\rm BV}$, only sufficiently strong magnetic fields 
become unstable and may escape to the outer layers of the star, whilst weaker 
magnetic fields are still unstable but with a much smaller growth rate, comparable
with the inverse lifetime of the star. 
For future work it will be needed to examine the case where the initial
field has a poloidal component, since in this case the Tayler instability may
behave differently \citep{dubama10}. 

It will also be necessary to explore the effects of rotation and shear:
we expect that, on one hand, rotation can stabilize the initial toroidal field,
whilst, on the other, shear may act as a source of 
the magnetic field. 
In this sense, the Tayler instability is known to be a symmetry breaking 
process, able to give rise to a saturated helical state starting from an 
infinitesimal helical perturbation \citep{BBDSM12}.
A possible dynamo effect might occur from the interplay between shear 
and Tayler instability. This processes needs to be investigated in the framework 
of global simulations used in this work. 
 
Finally, it is worth repeating the experiments with realistic stratification profiles
of A/B stars. For fiducial profiles of the gravity acceleration, the field might 
either remain confined to the interior or emerge towards the upper layers.  
In the latter case the use of open magnetic boundaries, i.e., vacuum or pseudo-vacuum, 
which could drive stellar winds, is necessary.  Mass loss through a stellar wind 
\citep{Alecian2019} as well as atomic diffusion \citep{Deal2016} are believed 
to play a significant role in explaining why magnetic A and B stars  
are chemically peculiar. 
Furthermore, there is a need to explore the dynamics of 
the magnetic field in the presence of thermohaline convection due to inhomogeneities 
in the plasma composition \citep[e.g.][]{Traxler+11}.
Having in hand more realistic models, we will be able to evaluate the
surface poloidal-to-toroidal field ratio as well as other diagnostics which could 
be directly compared to spectropolarimetric observations \citep[e.g.][]{Oksala2018MNRAS}.

\section*{Acknowledgments}
We thank the anonymous referee for his/her comments and suggestions. 
The work of F.D.S. has been performed under the Project HPC-EUROPA3 (INFRAIA-2016-1-730897), 
with the support of the EC Research Innovation Action under the H2020 Programme; in particular, 
G.G. and F.D.S. gratefully acknowledge the support and the hospitality of INAF Astrophysical 
Observatory  of Catania, and the computer resources and technical support provided by CINECA.


\bibliographystyle{mnras}
\bibliography{hh} 


\bsp	
\label{lastpage}
\end{document}